\documentclass[sigconf, nonacm]{acmart}

\usepackage[]{hyperref}

\usepackage{enumitem}
\usepackage{algorithm}
\usepackage[noend]{algpseudocode}
\usepackage{makecell}
\usepackage{multicol}
\usepackage{multirow}

\usepackage{adjustbox}
\usepackage[table]{xcolor}
\usepackage{tabularx}
\usepackage{booktabs}
\usepackage{array}     
\usepackage{xcolor}    
\usepackage{ragged2e}  

\newcolumntype{Y}{>{\Centering\arraybackslash}X} 
\newcolumntype{Z}{>{\RaggedRight\arraybackslash}X} 

\definecolor{ForestGreen}{RGB}{34,139,34}

\newcommand{\RN}[1]{%
  \textup{\uppercase\expandafter{\romannumeral#1}}%
}

\newlength\myindent
\newcommand{\THISWORK}{{\fontfamily{lmss}\selectfont
BlazingAML}}

\begin{document}
\title{\THISWORK: High-Throughput Anti-Money Laundering (AML) via Multi-Stage Graph Mining}

\author{Haojie Ye}
\affiliation{%
  \institution{University of Michigan}
}
\email{yehaojie@umich.edu}

\author{Arjun Laxman}
\affiliation{%
  \institution{University of Michigan}
}
\email{arlx@umich.edu}

\author{Yichao Yuan}
\affiliation{%
  \institution{University of Illinois Urbana-Champaign}
}
\email{yichaoy2@illinois.edu}

\author{Krisztian Flautner}
\affiliation{%
  \institution{University of Michigan}
}
\email{manowar@umich.edu}

\author{Nishil Talati}
\affiliation{%
  \institution{University of Illinois Urbana-Champaign}
}
\email{nishil@illinois.edu}

\begin{abstract}
Money laundering detection faces challenges due to excessive false positives and inadequate adaptation to sophisticated multi-stage schemes that exploit modern financial networks. 
Graph analytics and AI are promising tools, but they struggle with the fuzziness of laundering patterns, which exhibit structural and temporal variations. Conventional data mining techniques require the detailed enumeration of pattern variants, which not only complicates the analyst's task to specify them, but also leads to large run-time overheads and difficulty training accurate AI models.
We present \THISWORK, a scalable Anti Money Laundering (AML) system design that introduces a novel \textit{multi-stage framework} for expressing fuzzy money laundering patterns and a \textit{domain-specific compiler} that transforms high-level pattern descriptions into high-performance code for multiple hardware back-ends: CPU and GPU.
Our multi-stage abstraction decomposes complex laundering schemes into logical stages connected by graph operations, enabling diverse patterns to be expressed using unified primitives while capturing structural and temporal fuzziness. 
The compiler applies sophisticated optimizations, eliminating manual parallel programming requirements for financial analysts.
Evaluation in IBM AML data sets shows that \THISWORK\ achieves the same F1 score as state-of-the-art approaches while delivering a significant 210$\times$ and 333$\times$ higher speedup on CPU and GPU, with superior scalability.
\end{abstract}

\maketitle

\section{Introduction} \label{section:introduction}


Money laundering poses a severe threat to global financial stability, enabling organized crime and terrorism by disguising illicit proceeds~\cite{song2024identifyingmoneylaunderingsubgraphs, deprez2025networkanalyticsantimoneylaundering}.
Current Anti-Money Laundering (AML) systems rely on rule-based techniques that combine customer due diligence, transaction monitoring thresholds, and suspicious activity reporting to detect criminal activity~\cite{fan2024improving, fan2025deep, FATF2022}.
However, these conventional approaches face limitations: they generate excessive false positives, struggle to adapt to evolving laundering strategies, and fail to detect sophisticated multi-stage schemes exploiting modern financial networks~\cite{cardoso2022laundrograph, eddin2022antimoneylaunderingalertoptimization}.
Combining graph analytics and Artificial Intelligence (AI) offers a promising alternative that can identify complex patterns that traditional rule-based systems cannot capture. 

Graph and AI-driven techniques show promise for real-world money laundering detection~\cite{altman2024realistic,blanuvsa2024graph,fan2025deep,weber2018scalable,lin2024fraudgt} but face challenges due to the \textit{fuzzy} nature of laundering patterns.
Money laundering schemes have structural and temporal fuzziness~\cite{blanuvsa2024graph}: structural fuzziness involves varying numbers of intermediate accounts with the same topology, while temporal fuzziness allows flexible timing ordering between transactions.
These characteristics create two challenges for scalable AML systems: expressing fuzzy patterns and efficiently mining them at high throughput for real-time financial crime detection across massive transaction graphs.

Although there is research on exact pattern matching and subgraph mining~\cite{mackey2018chronological,paranjape2017motifs,talati2022mint,yuan2023everest}, little work has been done on systematically expressing fuzzy patterns or developing high-performance AML algorithms using them. Mining fuzzy patterns by reducing them to exact patterns causes a combinatorial explosion, which makes the approach of using exact mining for fuzzy patterns, at a minimum, impractical.

\begin{figure}
    \centering
    \includegraphics[width=0.48\textwidth]{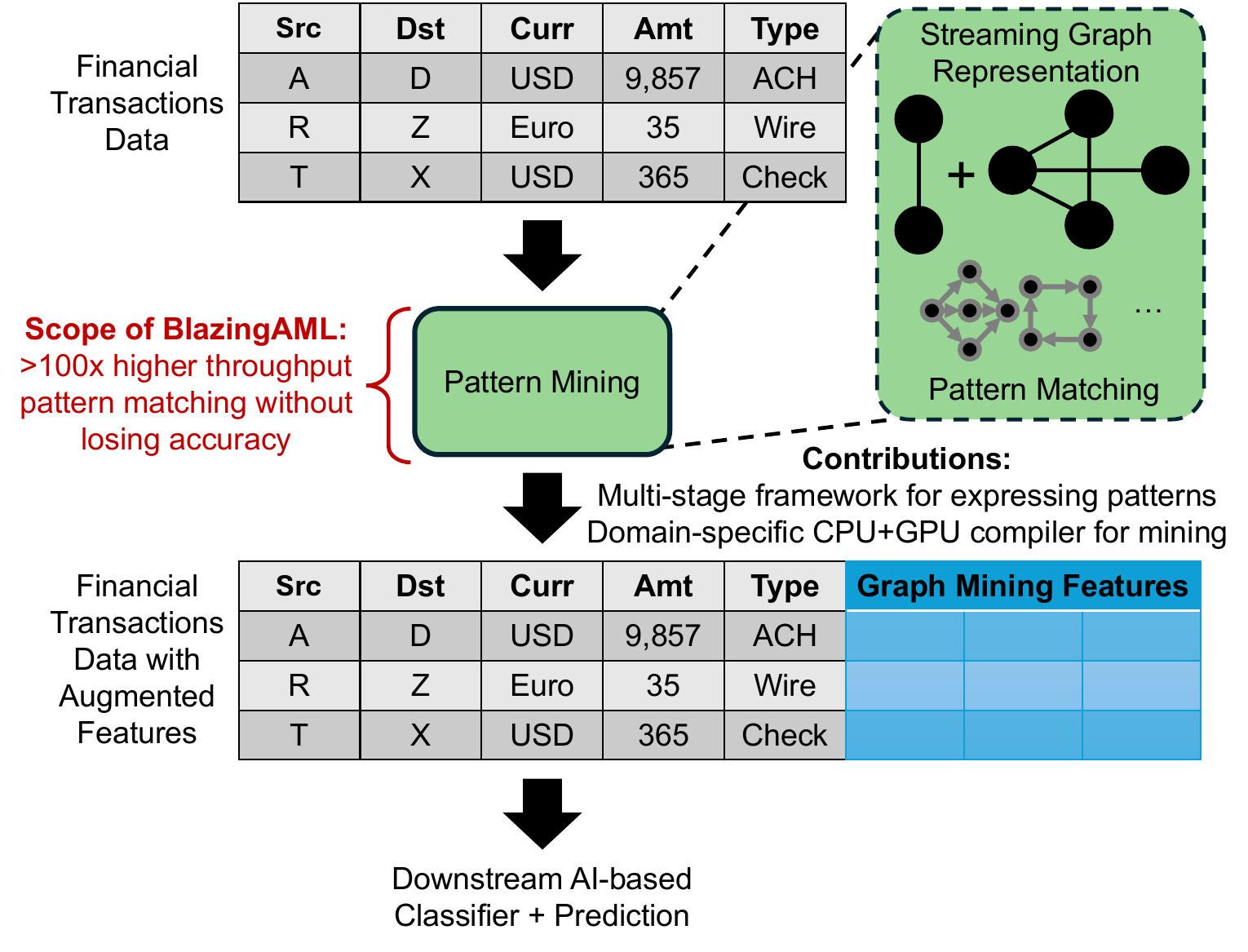}
    \caption{Overview and contributions of \THISWORK.}
    \label{fig:fig1}
\end{figure}

In this paper, we propose \THISWORK: a scalable AML system that systematically expresses fuzzy money laundering patterns and compiles them into high-performance mining code.
As shown in Fig.~\ref{fig:fig1}, our design follows a typical AML pipeline that combines graph pattern matching with AI techniques. 
The first stage employs specialized graph mining to detect known laundering patterns, such as scatter-gather schemes and multi-hop cycles. It augments each transaction edge with pattern occurrence counts as additional features.
A downstream gradient boosting classifier~\cite{chen2016xgboost} then leverages these enriched features to classify individual transactions as legitimate or illicit.
While a more sophisticated AI-based classifier (\textit{e.g.,} graph neural network~\cite{lo2023inspection, johannessen2023findingmoneylaunderersusing}) can improve AML accuracy, this paper focuses on improving graph pattern mining expressibility and throughput.
Trying new classifiers is a compelling direction for future work.

The \textit{main contribution of this work is the graph mining pipeline} optimized for AML:
\begin{enumerate}[leftmargin=*]
\item A \textit{multi-stage specification} technique that captures structural and temporal fuzziness in laundering patterns, enabling analysts to describe variable-topology schemes and flexible timing constraints in a unified framework.
\item A \textit{domain-specific compiler} that transforms high-level pattern descriptions into optimized C++ and CUDA kernels, handling graph-specific optimizations.
\end{enumerate}

Our multi-stage specification technique introduces a unified abstraction that decomposes complex money laundering patterns into logical stages connected by fundamental graph operations.
This allows diverse laundering schemes to be expressed using the same high-level primitives while naturally capturing both structural and temporal fuzziness inherent in real-world patterns.

This approach establishes a description language for AML patterns that allows domain experts to specify only the logical structure of detection algorithms while automatically generating optimized implementations for different hardware architectures, effectively separating pattern logic from performance optimization concerns. 

The framework's modular design enables scalable pattern mining on graphs with millions of nodes through automatic parallelization and incremental processing, while supporting extensibility through simple pattern library modifications rather than fundamental algorithmic reimplementation.

The specifications are compiled by a domain-specific compiler from high-level declarative pattern specifications into optimized C++ and CUDA kernels. The optimizations performed by the compiler include:

\begin{itemize}[leftmargin=*]
\item Power-law-aware memory access pattern generation,
\item Degree-based workload balancing, and
\item Pipelined CPU-GPU execution based on pattern structure.
\end{itemize}

Our system enables rapid development, specification, and deployment of optimized fuzzy graph mining algorithms by providing a seamless flow from specification to optimized CPU-GPU code. Combined with a gradient-boosted classifier, we can create an end-to-end AML pipeline, forming a system we refer to as \THISWORK.

To evaluate the effectiveness of \THISWORK, we compare both accuracy and speed using AML datasets released by IBM~\cite{altman2024realistic}.
Our evaluation demonstrates that \THISWORK\ achieves significantly higher throughput.
By mining graph patterns and using them as augmented features, identical in value to those used by GFP~\cite{blanuvsa2024graph}, \THISWORK\ attains the same level of F1 score as GFP while being substantially faster.

In particular, our experiments show average speedups of 210$\times$ on CPU and 333$\times$ on GPU compared to GFP.
Moreover, \THISWORK\ exhibits superior scalability, maintaining performance advantages as the size of the input transaction graph increases.
We further benchmark against a graph transformer–based approach, FraudGT~\cite{lin2024fraudgt}. 
While FraudGT achieves a higher F1 score through computationally expensive model training and inference, \THISWORK\ delivers 4.9$\times$ higher throughput, making it far more practical for large-scale, real-time AML workloads.
\THISWORK\ makes the following novel contributions.

\begin{itemize}[leftmargin=*]
    \item Design of a \textit{multi-stage framework} to flexibly express money laundering patterns.
    \item Design of a \textit{compiler} that outputs pattern-specific high-performance code for multiple hardware back-ends.
    \item \THISWORK: an \textit{end-to-end AML system design} that delivers 4.9$\times$ higher throughput compared to the state-of-the-art with the same level of accuracy.
\end{itemize}

\section{Background and Related Work} \label{section:background}

\subsection{Money Laundering} \label{subsec:background_ml}
Money laundering is the process by which criminals disguise the illicit origin of funds to integrate them into the legitimate financial system.
If undetected, it fuels organized crime, enabling drug cartels, human trafficking rings, and terrorist organizations that cause immense human suffering.
For example 150,000 murders are linked to Mexican drug cartels since 2006 and an estimated 40 million people have been enslaved through trafficking~\cite{weber2018scalable}.
The UN estimates up to 5\% of global GDP, or roughly 2 trillion USD, is laundered each year, with global financial crimes totaling 3.5 trillion USD in 2020~\cite{ey}.
The fast growth of digital transaction businesses also gives rise to more sophisticated financial crimes. 

Money laundering typically unfolds in three stages: (1) \textbf{Placement}, where illicit proceeds are broken into smaller deposits to avoid detection~\cite{schneider2004money},
(2) \textbf{Layering}, involving complex fund transfers among shell companies and accounts to obscure origins, and (3) \textbf{Integration}, where cleaned funds are reintroduced into the economy through assets like real estate or securities.
Money laundering detection is primarily focused on the \textbf{Layering} stage, where illicit funds are circulated through the financial system in various forms and transactions to obscure their origin.
Despite tens of billions of dollars spent annually on compliance~\cite{kpmg2014}, and severe penalties such as the 530 million USD fine levied on the Commonwealth Bank of Australia in 2018~\cite{bbc2018}, Europol estimates only 1\% of illicit funds are recovered~\cite{europol2017}.

Detecting money laundering, called Anti Money Laundering (AML), is an extremely challenging technical problem.
It is a \textit{needle-in-a-haystack} problem in massive, noisy, fragmented, and constantly changing transaction datasets, often spread across multiple institutions and jurisdictions~\cite{weber2018scalable}.
Below, we outline the current real-world practices of AML and present a set of advanced graph-based techniques designed to automate detection and achieve scalable AML monitoring.

\subsection{Rule-Based AML} \label{subsec:background_ruleaml}
Today, financial institutions primarily rely on rule-based techniques to detect and prevent money laundering. These approaches combine regulatory compliance requirements, structured monitoring, and risk assessment procedures to identify suspicious behavior, flag potential illicit activity, and ensure adherence to AML standards. The following sections outline key components of conventional AML practices~\cite{fan2025deep}.

\begin{itemize}[leftmargin=*]
    \item \textbf{Customer due diligence.} Financial institutions must verify the identities of their clients through comprehensive know-your-customer (KYC) procedures. This process entails gathering personal identification information, performing background checks, and evaluating risk profiles based on factors such as geographic location, transaction patterns, and credit history.

    \item \textbf{Transaction monitoring.} Continuous surveillance of customer transactions is essential to detect irregular or suspicious activity. This includes monitoring for unusually large deposits, structuring of transactions to avoid reporting thresholds, and transfers involving high-risk individuals, entities, or jurisdictions.

    \item \textbf{Suspicious activity reporting (SAR).} When potential money laundering is identified, institutions are required to submit SARs to local financial intelligence units~\cite{fan2024improving}. These reports enable authorities to investigate suspicious transactions and support law enforcement in taking appropriate action.

    \item \textbf{Internal audits and compliance programs.} Regular internal audits are necessary to ensure AML programs remain compliant with evolving regulations, address operational gaps, and counter increasingly sophisticated laundering methods~\cite{FATF2022}. These audits are vital for evaluating the effectiveness of existing controls and identifying vulnerabilities, particularly given the complexity and scale of transactions facilitated by digital and mobile platforms.

\end{itemize}

Rule-based techniques for combating money laundering struggle to keep up with the increased sophistication of financial transactions, particularly due to the widespread adoption of mobile payments and interconnected networks.
This complexity leads to high false-positive rates, burdening AML teams and allowing criminal activities to go undetected, underscoring the need for advanced detection mechanisms.

\subsection{Graph and AI-based AML} \label{subsec:background_graphdl_aml}
Financial transaction data can be naturally represented as graphs, where a node represents a bank account and an edge between a pair of nodes represents a financial transaction between two accounts.
Attributes on nodes and edges represent the amount of money transferred, currency, bank account details, etc.
Converting a financial transaction database into a graph representation presents an opportunity to use graph analytics, AI, or a combination of these two techniques for detecting money laundering.
Below, we discuss three broad categories of techniques used for AML.

\begin{figure}
    \centering
    \includegraphics[width=0.48\textwidth]{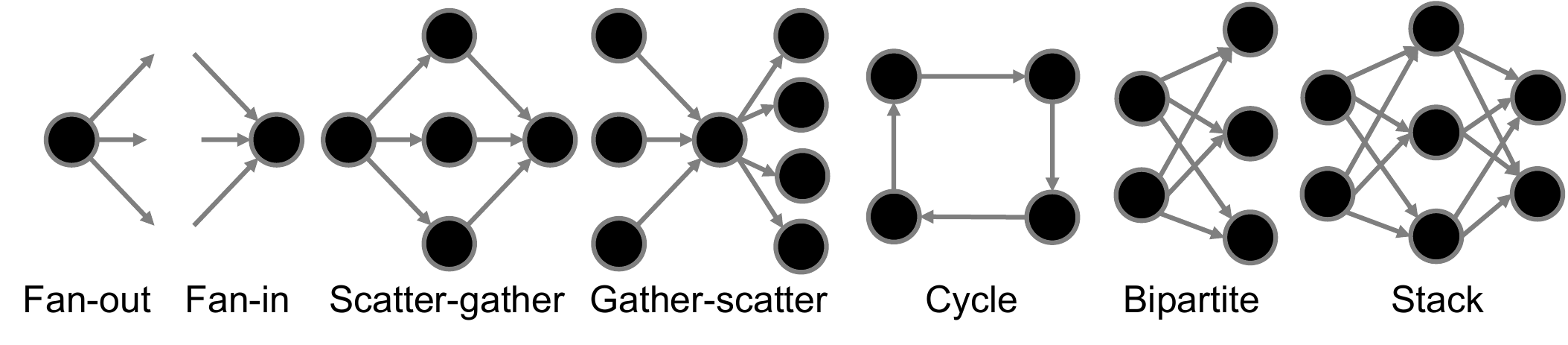}
    \caption{Representative graph patterns illustrating layering strategies in money laundering within financial transaction graphs~\cite{altman2024realistic}.}
    \label{fig:money_base_patterns}
\end{figure}



\begin{itemize}[leftmargin=*]
    \item \textbf{Graph pattern matching-based techniques.} 
    Graph pattern matching methods identify suspicious motifs in transaction graphs: such as cycles, cliques, or hub–spoke topologies~\cite{altman2024realistic,blanuvsa2024graph} (see Fig.~\ref{fig:money_base_patterns}).
    These structural patterns are often linked to established money laundering strategies, including circular layering and smurfing~\cite{weber2018scalable}.
    More recent approaches extend this idea by employing subgraph mining and filtering techniques to expose anomalous relational structures that may signal laundering flows~\cite{egressy2024provably}.

    \item \textbf{AI-based techniques.} 
    Pure AI methods represent transactions as sequences or features and apply neural models for classification or anomaly detection. 
    CNNs can extract spatial-temporal correlations in transaction matrices~\cite{singh2021machine}; RNNs such as GRU/LSTM capture sequential dependencies in mobile or banking transaction streams~\cite{bian2022machine}; 
    transformers model long-range dependencies across transaction histories~\cite{tatulli2023hamlet, zhang2024generative}; 
    and deep reinforcement learning has been explored by formulating AML detection as a sequential decision task~\cite{uprety2020reinforcement}. 
    These methods can adapt to evolving laundering strategies, though they often face challenges of imbalance and limited real-world labels.

    \item \textbf{Hybrid graph and AI-based techniques.}
    Hybrid methods leverage both structural and learned features. 
    For example, Inspection-L combines self-supervised graph embeddings with a downstream Random Forest classifier~\cite{lo2023inspection}. 
    Graph neural networks and their extensions directly capture relational dependencies, with methods such as EvolveGCN~\cite{pareja2020evolvegcn}, LaundroGraph~\cite{cardoso2022laundrograph}, FraudGT~\cite{lin2024fraudgt}, and GAGNN~\cite{cheng2023anti}.
    GFP~\cite{blanuvsa2024graph} incorporates pattern counts as edge features, which are then provided to a lightweight downstream classifier (\textit{e.g.,} XGBoost~\cite{chen2016xgboost}).
    These approaches enrich traditional AI models with graph context, leading to state-of-the-art performance on datasets like Elliptic~\cite{weber2019anti} and IT-AML~\cite{altman2024realistic}.
\end{itemize}

A key challenge with purely graph pattern matching–based techniques is that money laundering strategies continuously evolve, making static motif detection insufficient. 
Moreover, as highlighted in prior work~\cite{altman2024realistic,blanuvsa2024graph}, the presence of a particular pattern does not necessarily indicate fraudulent activity. 
For instance, fan-in and fan-out structures are common in legitimate account activity and are therefore not inherently suspicious.
Conversely, purely AI-based approaches often struggle to capture the rich structural interactions between nodes and edges in transaction graphs.
To address these limitations, this work proposes a hybrid system that integrates graph-based and AI-based methods.
Following the design philosophy of GFP~\cite{blanuvsa2024graph}, our approach employs graph pattern matching in the front end to extract known suspicious patterns (see Fig.~\ref{fig:money_base_patterns}).
The counts of these patterns are then incorporated as node and edge features, which are subsequently used by a downstream AI classifier to detect potential money laundering behavior.

\section{Challenges in AML System Design} \label{sec:challenges}

This section discusses the challenges of designing a scalable detection system for AML.

\textbf{High data volumes.} 
Detecting money laundering is particularly challenging due to the immense volume of financial transactions that must be monitored. The proliferation of digital and mobile payment platforms, as well as smart IoT devices, has further exacerbated this problem by dramatically increasing both the scale and velocity of transactional data~\cite{fan2025deep}. These systems generate millions of heterogeneous records daily, often across multiple payment channels, making rapid detection increasingly complex.
Moreover, criminals exploit this high-volume environment by structuring illicit flows into smaller, inconspicuous digital transactions (\textit{e.g.,} smurfing), which are easily obscured within legitimate traffic~\cite{fan2025deep}.
Consequently, modern AML systems must not only scale to vast transaction streams but also remain adaptive to the evolving complexity of digital ecosystems.


\textbf{High-throughput requirement.} 
An effective AML system must achieve high throughput to process massive transaction streams in real time and prevent illicit funds from being integrated into the financial system. The ability to sweep through large volumes of data quickly is critical, as delays in detection can allow criminals to move assets across borders or convert them into untraceable forms, severely limiting recovery and enforcement efforts. However, designing high-throughput AML systems is challenging due to the need to balance scalability with accuracy; processing data at speed often increases false positives and risks overlooking subtle laundering patterns. Furthermore, heterogeneous data sources and evolving transaction structures demand architectures that are both computationally efficient and adaptable to new laundering strategies. 

\textbf{Complex transaction patterns.} 
As shown in Fig.~\ref{fig:money_base_patterns}, graph pattern matching in AML requires mining highly complex structures, such as cycles, cliques, and multi-hop motifs—that capture subtle laundering behaviors~\cite{blanuvsa2024graph}.
Algorithms that can identify these patterns at scale is computationally expensive, as the search space grows combinatorially with the size of the pattern and transaction graph.
This makes it particularly challenging to achieve high throughput when processing massive financial datasets, where billions of edges must be examined for rapid detection.
Consequently, as shown in our results, scalable graph mining for AML remains a core bottleneck in building efficient detection systems.


\begin{figure}
    \centering
    \includegraphics[width=0.48\textwidth]{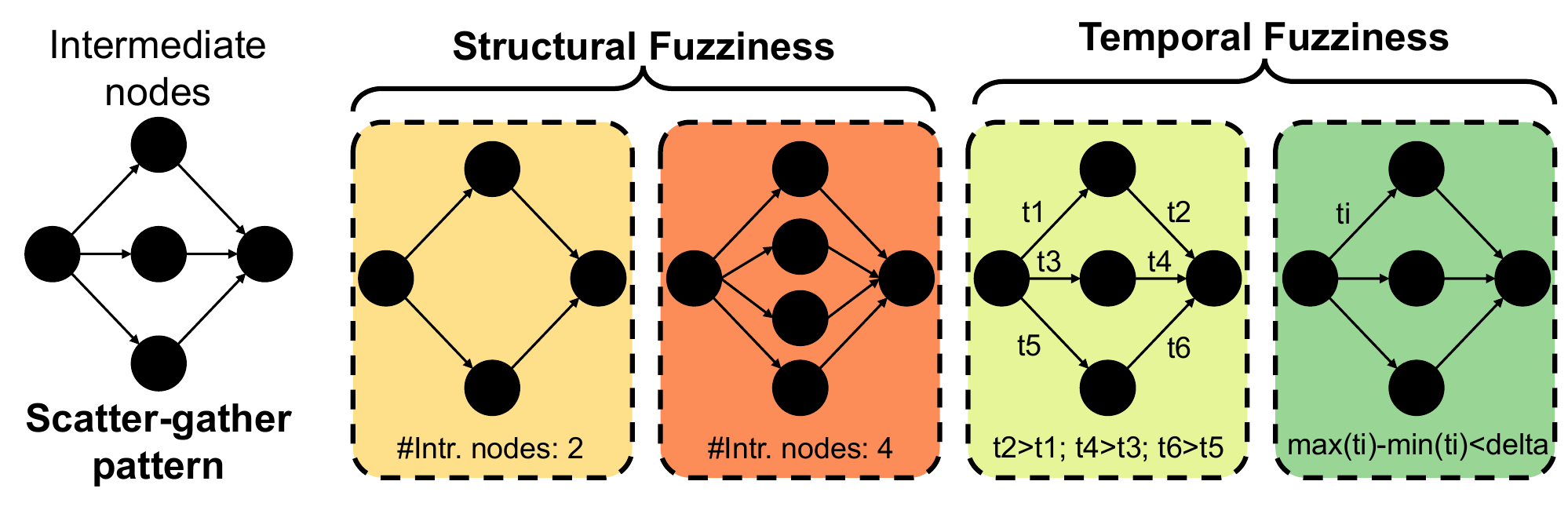}
    \caption{Fuzziness illustrated in the scatter-gather money laundering pattern: (1) structural fuzziness in terms of different numbers of intermediate nodes, and (2) temporal fuzziness in terms of partial time ordering and time window.}
    \label{fig:scatter_gather_fuzziness}
\end{figure}
\textbf{Fuzzy patterns.}
While there is a large body of work designing algorithms that work for mining a fixed pattern, money laundering patterns exhibit fuzziness across two axes, as shown in Fig.~\ref{fig:scatter_gather_fuzziness} as detailed below.
\begin{itemize}[leftmargin=*]
    \item \emph{Structural fuzziness:} Although a pattern has a fixed structural shape, the number of nodes and edges involved can vary significantly.
    For instance, in a scatter–gather pattern, the number of intermediate \textit{placement} accounts between the source and destination may range from just a few to dozens.
    Laundering actors can freely choose the number of intermediaries and the connectivity structure to evade detection. 
    Enumerating and matching all such variants would require mining a combinatorial number of exact patterns, which is computationally expensive.
    \item \emph{Temporal fuzziness:} 
    While many temporal mining frameworks~\cite{paranjape2017motifs,mackey2018chronological,yuan2023everest} impose a strict global temporal ordering of edges within a time window, money laundering patterns often do not adhere to such rigid structures.
    In practice, transactions may not follow a global sequence.
    For example, the scatter and gather phases can be temporally decoupled, with only local or partial ordering constraints (\textit{e.g.,} $t_i < t_{i+1}$) holding.
    As a result, even minor deviations in temporal order can cause overly rigid pattern-matching systems to miss true positives.
\end{itemize}

Structural and temporal fuzziness is challenging to capture for a neural network, given the amount of training data that's available. The "XGB Only" column of Table~\ref{table:f1xgb} illustrates the low F1 scores without graph-pattern-derived features. The rightmost column in the table shows the significantly better results after the graph has been annotated with the mined patterns, thereby supporting the argument that pattern-mining should be used to generate features that increase model accuracy.

Existing commercial graph query languages, like Cypher \cite{cypher}, are limited in their ability to express the structural and temporal fuzziness of money laundering patterns.
Cypher can handle basic fuzzy structures like variable-length paths, but it lacks syntax to express complex topological variations within a single query.
Moreover, Cypher lacks support for temporal fuzziness, where edge orderings may follow partial temporal constraints and time windows may overlap.
This forces analysts to decompose fuzzy patterns into rigid subqueries, losing the holistic view necessary for effective pattern mining and requiring complex post-processing. 

\textcolor{black}{
Recent graph-mining systems have introduced “anti-edge’’ and “anti-vertex’’ constructs to express absence constraints in pattern queries. 
For example, Peregrine~\cite{jamshidi2020peregrine} explicitly provides anti-edge/anti-vertex operators to forbid certain vertices or edges from matching, and~\cite{jamshidi2022anti} defines an anti-vertex to exclude specific neighbors in subgraph queries declaratively. 
These constructs are designed for general graph-pattern tasks (e.g., filtering maximal cliques or anomalies by excluding specific \textbf{single} or \textbf{a few} unwanted connections, typically enumerated manually by the authors), emphasizing expressive pattern constraints rather than any AML-specific logic. 
By contrast, \THISWORK\ targets fundamentally different challenges: analyzing massive, noisy transaction graphs with inherently “fuzzy’’ structural and temporal patterns. AML patterns do not follow fixed anti-vertex structures but instead describe the procedural logic of how money flows in a laundering strategy. 
Effective AML detection must tolerate both structural and temporal ambiguity (e.g., \textbf{any number} of approximate or missing edges and flexible time windows), which rigid anti-edge/anti-vertex semantics cannot capture. 
In short, while anti-edge/anti-vertex constructs are conceptually interesting, they address a distinct problem space and do not overlap with \THISWORK\ ’s core contributions in automated, high-performance, and scalable anti-money-laundering pattern mining over high-volume real-time financial transactions. 
}

\textbf{Distinguishing between innocent vs. fraudulent transactions.} 
Many transaction patterns also appear in benign scenarios, necessitating joint reasoning over structural, temporal, and contextual features to reduce false positives.

While rule-based and purely AI-driven AML approaches provide partial solutions, the most effective approaches increasingly adopt a hybrid paradigm~\cite{blanuvsa2024graph} that integrates graph pattern mining with machine learning classifiers.
This combination captures structural irregularities and adaptive behavioral cues, improving robustness against evolving laundering strategies.
However, existing pattern mining systems~\cite{blanuvsa2024graph,altman2024realistic,yuan2023everest,mackey2018chronological,paranjape2017motifs} lack expressive frameworks to flexibly model real-world laundering patterns, especially those with fuzzy structural or temporal variations.
They also fail to scale when applied to massive, streaming transaction graphs required for practical AML deployment.
For instance, the IBM system~\cite{blanuvsa2024graph}, despite strong detection accuracy, demonstrates a steep performance decline as transaction volumes increase (as shown in our evaluation), highlighting the limitations of current graph engines in high-throughput settings.
These gaps motivate the \THISWORK\ system design, which unifies flexible pattern specification with scalable execution to realize hybrid AML detection at a practical scale.

\section{\THISWORK\ Design Goals}
\label{sec:why_compiler}


Prior graph mining systems ~\cite{paranjape2017motifs,yuan2023everest,mackey2018chronological} assume fixed shapes and strict edge orders for patterns that are infeasible to employ in a large-scale AML setting due to pattern fuzziness.

Detecting scatter-gather money laundering patterns with rigid pattern definitions for each variant leads to multiple distinct algorithm implementations with different graph traversal logic and temporal validation rules. The computational complexity becomes prohibitive as pattern variations increase: for a 3-size scatter-gather pattern, existing frameworks must enumerate $6! = 720$ distinct temporal constraint combinations, making exhaustive enumeration computationally infeasible for large-scale or real-time detection systems.
Even with identical structural topology, exact algorithms must enumerate all possible combinations of partial temporal constraints, creating a combinatorial explosion that scales as $O(n!)$ where $n$ is the number of participating edges. This factorial growth renders traditional approaches impractical for realistic money laundering scenarios.

Furthermore, each exact algorithm must be separately optimized and maintained while redundantly scanning the same graph regions multiple times, resulting in enormous implementation complexity and computational overhead. The proliferation of specialized algorithms creates a maintenance burden that grows quadratically with the number of supported pattern variants.
To address these fundamental limitations of exact pattern matching, we designed a unified fuzzy pattern matching compiler with the goals discussed below, which captures all structural and temporal variations in a single multi-stage algorithm.

\textbf{Expressivity beyond fixed templates.}
Analysts need to articulate domain rules that go beyond rigid motifs. 
For instance, in the scatter--gather (smurfing) and cycled laundering scenario, a bank may require that a lower bound on the number of placement edges (\textit{e.g.,} ``at least N''), but not mining a specific size, as shown in Fig.~\ref{fig:scatter_gather_fuzziness}.

\textbf{Flexible temporal semantics.}
Financial behaviors frequently obey only partial orders and window constraints: ``integration occurs after placement'' within a time horizon $\delta$, but events inside each phase are mutually interchangeable. 
Similarly, in cycled laundering, funds may traverse a cycle with edges that are \emph{not} observed in strict cycle order (\textit{e.g.,} account~A credits B earlier as camouflage, before illicit funds arrive at~A).
Fixed-shape, exact-order miners cannot directly express these variants without enumerating all permutations.
A practical pattern language must (i) represent steps as \emph{logical time advances} with optional per-step partial orders; (ii) allow cross-step constraints (\textit{e.g.,} causality, min/max fanout, node inequality); and (iii) support out-of-order evidence (anticipatory edges) provided eventual consistency constraints are satisfied. 
Prior frameworks that pin every edge to a global strict order miss these semantics.

\textbf{Separation of concerns for analysts.}
Most AML users are not graph systems experts. 
They should specify \emph{what} constitutes suspicious behavior, not \emph{how} to implement neighbor enumeration, set intersections, or pruning strategies. 
A flexible compiler decouples high-level intent from low-level execution, enabling non-experts to encode complex constraints safely and audibly.
The compiler can map declarative stages to optimized kernels and loops (neighbor ordering, early exits on temporal violations, degree-aware intersections, workload balancing across threads/warps/cores), and target multiple backends (CUDA for GPUs, OpenMP for CPUs). 
This preserves scalability on power-law transaction graphs while freeing analysts from hand-tuning.

\textbf{Why \THISWORK?}
These requirements directly motivate our design in \S\ref{sec:pattern_expression}. We introduce \emph{Temporal Segment Composition}, which represents patterns as sequences of logical time steps with interchangeable operations inside a step and explicit cross-step constraints. This abstraction simultaneously captures structural and temporal fuzziness while remaining compiler-friendly: it admits static checks, enables aggressive pruning, and maps cleanly to specialized CUDA and OpenMP code. In the next section, we detail this representation and how it underpins our code generation pipeline.

\section{Multi-Stage Framework for Expressing Money Laundering Graph Patterns} \label{sec:pattern_expression}

A central challenge in detecting money laundering lies in the diversity and structural complexity of illicit transaction patterns, as discussed in \S\ref{sec:challenges}.
To address this, we present a novel \textbf{multi-stage framework} for expressing and detecting money laundering patterns in large-scale financial transaction graphs.
The core innovation lies in decomposing complex laundering patterns into a series of logical stages, each representing a distinct phase in the money laundering process.
A \textbf{\textit{stage}} captures how money flows from one entity to another or how previously discovered transaction chains can be systematically extended.
Each stage is produced by applying specific \textbf{\textit{operations}} (such as neighbor expansion, set intersection, or union) on one or multiple previous stages.
These operations act on fundamental graph elements called \textbf{\textit{operands}}, which can be nodes, edges, or the outputs from preceding stages.
By chaining these stages together, analysts can describe sophisticated laundering patterns in a systematic and computationally tractable manner.

\begin{figure}
    \centering
    \includegraphics[width=0.48\textwidth]{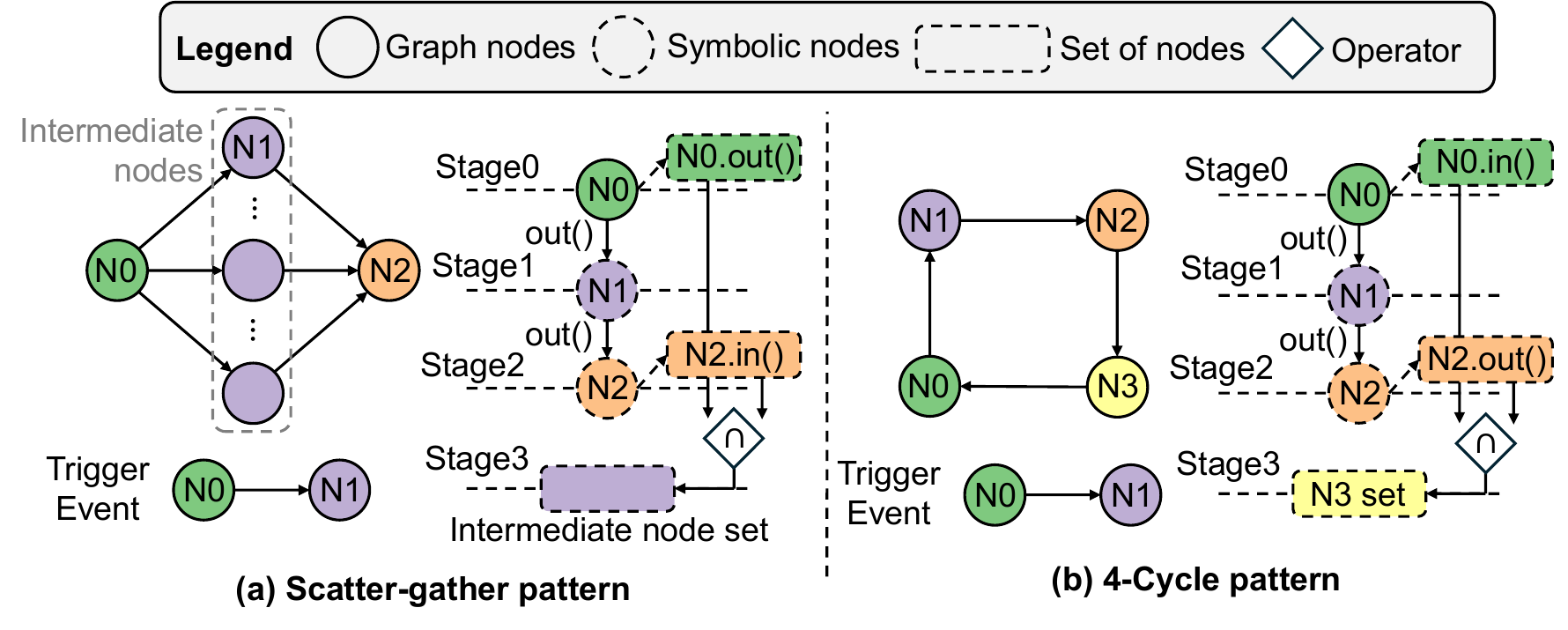}
    \caption{Example of (a) scatter-gather and (b) 4-cycle patterns expressed in the proposed multi-stage framework. Note the same stage-based structure for the two patterns, with different in- and out-neighbor sets at stages 0 and 2.}
    \label{fig:multi_stage_pattern_example}
\end{figure}

\textbf{Illustrative examples: scatter-gather and cycle patterns.}
Consider the \textbf{scatter-gather pattern}, a common technique where funds are dispersed through multiple accounts before being reconsolidated.
In our proposed multi-stage framework, this pattern is decomposed as follows as shown in Fig.~\ref{fig:multi_stage_pattern_example}(a): Stage 0 to Stage 1 represents the initial trigger transaction from node N0 to N1, initiating the mining process.
Stage 1 to Stage 2 involves traversing the out-neighbors of N1 (denoted as N2), effectively capturing potential layering accounts where funds are dispersed.
Stage 2 to Stage 3 performs an intersection operation between N2's in-neighbors and N0's out-neighbors, identifying candidate gathering accounts where funds reconverge.
Symbolic nodes in the figure represent either a single graph node or a node set within each stage.

Notably, multiple intermediate accounts may be discovered in parallel, which naturally captures \textit{structural fuzziness} in laundering patterns (Fig.~\ref{fig:scatter_gather_fuzziness}).
Beyond structure, \textit{temporal fuzziness}, not shown in Fig.~\ref{fig:multi_stage_pattern_example} for simplicity, can also be expressed by introducing temporal ordering or window constraints as needed.
For instance, a temporal ordering between outgoing and incoming edges of intermediate nodes (\textit{i.e.,} the first option illustrated under temporal fuzziness in Fig.~\ref{fig:scatter_gather_fuzziness}) can be encoded by adding an ordering constraint between edges spanning Stage 2 to Stage 1 and Stage 1 to Stage 0.

Similarly, \textbf{cycle} detection, another fundamental laundering structure, can be expressed using identical primitives as shown in Fig.~\ref{fig:multi_stage_pattern_example}(b).
The cycle pattern begins with Stage 0 to Stage 1 representing the trigger transaction N0 to N1, followed by Stage 1 to Stage 2 involving expansion to N2.
Finally, Stage 2 to Stage 3 performs an intersection of N2's out-neighbors with N0's in-neighbors, effectively closing the loop to yield a complete cycle.
Despite being structurally different from scatter-gather patterns, cycles are expressed using the same basic operations: neighborhood expansion and intersection, highlighting the generality and unifying power of our proposed framework.

\textbf{Integration with streaming analytics and machine learning.}
Real-world financial transactions arrive in a streaming fashion, making it crucial to design mining algorithms that can operate over continuously arriving edges in financial transaction graphs.
Our framework seamlessly integrates with modern streaming analytics and machine learning pipelines.
Each incoming transaction edge updates a time-windowed feature representation, while the mining process maintains comprehensive feature lists across all pattern instances. 
When a new edge arrives in the transaction stream, it automatically increments counts for all pattern instances it participates in, such as scatter-gather patterns of various sizes or different cycle configurations. 
These continuously updated feature vectors are then fed into machine learning models such as XGBoost~\cite{chen2016xgboost}, enabling a hybrid approach that combines rule-based pattern detection with statistical learning.
This tight integration ensures that the framework transcends traditional symbolic pattern matching, supporting sophisticated feature-driven decision-making that can adapt to evolving laundering techniques.

\textbf{Rationale and advantages of the multi-stage framework.}
The decomposition of money laundering detection into discrete stages provides several compelling advantages that address fundamental challenges in AML.
\begin{enumerate}[leftmargin=*]
\item The stage abstraction naturally captures the logical flow of money laundering operations, with each stage representing a unique transaction flow in the network.
This alignment mirrors established financial investigation practices, making the framework intuitive for domain experts. 
\item The modular nature of stages enables analysts to construct complex patterns from simple, reusable building blocks, promoting both code reuse and pattern library development.
\item The framework inherently supports parallelism and node interchangeability: operations within a stage can usually be executed simultaneously across multiple candidate nodes, enabling efficient utilization of modern GPU and multi-core CPU architectures.
\end{enumerate}

The proposed approach introduces several novel dimensions to AML research.
First, the framework provides a \textit{unified abstraction} where diverse laundering patterns, including scatter-gather, cycles, and chain structures, can all be expressed using the same high-level primitives.
This standardization establishes a common descriptive language for AML patterns, facilitating pattern library development and cross-institutional benchmarking. 
Furthermore, the proposed abstraction allows analysts to specify only the logical structure of patterns while the system automatically determines optimal backend implementations (as discussed in the next section), whether using OpenMP for CPU parallelization, CUDA for GPU acceleration, or specialized graph-optimized routines.
Complex optimization decisions, such as exploiting power-law graph structures or dynamically selecting smaller neighborhoods at runtime, are entirely encapsulated within the compiler, hiding implementation complexity from domain experts such as bankers and analysts.

The framework's design directly addresses the scalability challenges inherent in real-world financial crime detection.
By leveraging automatic parallelization and stage-wise decomposition, the system scales effectively to financial graphs containing millions of nodes and edges, representative of major financial institutions' transaction volumes.
The modular stage architecture also enables incremental processing, where new transactions can trigger localized pattern updates rather than requiring full graph recomputation.
Additionally, the extensibility of the framework means that new laundering patterns can be incorporated by simply modifying stage definitions, without requiring fundamental algorithmic re-implementation.
This combination of performance optimization, scalability, and adaptability positions the multi-stage framework as a significant advancement in computational approaches to financial crime detection, offering both theoretical elegance and practical utility for large-scale anti-money laundering systems.

\section{A Domain-Specific Compiler for Multi-Stage AML Pattern Mining} \label{sec:compiler}

This section presents the design of a compiler that takes multi-stage AML pattern description (\S\ref{sec:pattern_expression}) and outputs high-performance code with CPU and GPU back-ends.

\textbf{Design goals and architecture.}
Our system introduces a domain-specific compiler that transforms high-level declarative AML pattern specifications into highly optimized C++ and CUDA kernels.
The primary design goal is to enable AML analysts to focus purely on the logical description of suspicious transaction patterns, without requiring any furtner manual generation and optimization of complex parallel graph-processing code.
This abstraction layer bridges the gap between domain expertise in financial crime detection and the technical complexity of high-performance graph computing.

The compiler accepts declarative input specifications that contain sequences of logical stages describing target patterns in terms of fundamental graph traversal primitives.
These primitives include \textbf{\texttt{for\_all}} operations for iteration over all edges or neighbors of a node, \textbf{\texttt{intersection}} operations for simultaneous matching of neighbor sets between nodes, \textbf{\texttt{differentiate}} operations for filtering with conditional logic such as eliminating self-connections, and \textbf{\texttt{operands}} representing graph entities including nodes, edges, and their attributes like source, destination, and transaction timestamps. 
Input files (\textit{e.g.,} YAML configurations) serve as one example format for expressing these specifications, though the compiler architecture supports multiple input representations.

\textbf{\textcolor{black}{Compilation of any general patterns.}}
\textcolor{black}{
The fundamental uniqueness of \THISWORK’s framework is that it allows users to express any general AML pattern as a logical procedure, rather than forcing them into a rigid motif shape defined by a fixed sequence of temporal edges. 
In our formulation, the analyst specifies the desired laundering logic using a set-operation language that precisely describes how money laundering entities evolve over the course of the laundering pattern. 
This design isolates the user’s conceptual pattern definition from the details of \THISWORK’s optimized execution flow, which must account for graph size, power-law distributions, hardware heterogeneity (CPU/GPU), and production deployment concerns such as data movement and memory layout.
}

\textcolor{black}{
Under this abstraction, \THISWORK\ defines any mined pattern as a composition of set operations, embedding within these operations all forms of structural and temporal fuzziness that naturally arise in AML investigations. 
To support this, \THISWORK\ breaks down the logical flow of a laundering scheme into a sequence of stages. 
Each stage represents a logical advancement of the flow—e.g., layering, scattering, reconvergence—rather than a single observed transaction. A stage declares its input set, output set, and the operation that transforms one into the other, along with node constraints (e.g., account type, currency type) and temporal constraints (e.g., time window or ordering).
}

\textcolor{black}{
The set operation at a stage can be exact—enumerating all neighbors and thus matching the semantics of traditional temporal motifs~\cite{paranjape2017motifs}—or it can be defined as any expression over sets emitted by earlier stages, such as union, intersection, difference, or general set differentiation. 
This flexibility lets the user express both tightly constrained patterns and highly fuzzy behaviors where multiple branches of any number of accounts or non-strict temporal orderings are allowed. 
The input sets for a stage may be newly defined or may reference any prior stage’s outputs in any time order, enabling broad expressive power. 
In contrast, classical temporal motifs restrict each edge to depend solely on the immediately previous edge and require monotonically increasing timestamps; this becomes just a special case of \THISWORK’s general template.
}

\textcolor{black}{
By filling in this template, an analyst can describe an arbitrary AML pattern—simple or complex—using concise logical steps. Once the pattern is specified, \THISWORK\ converts each stage into optimized CPU/GPU code independently. Each stage is compiled into a nested-loop structure: the compiler generates an outer loop over the stage’s input set, and inner logic that performs the required set operations using operands drawn either from freshly constructed sets or from any previously materialized stage outputs. 
The compiler tracks data dependencies to determine whether a stage’s results should feed forward into later stages or whether an earlier stage must be revisited to supply the operands needed by a downstream set operation (e.g., to fulfill a required \texttt{union}, \texttt{intersection}, \texttt{difference}, or \texttt{differentiate} over future stages).
}

\textcolor{black}{
To make the compilation process more concrete, Algorithm~\ref{alg:nestedloopcell} shows the generic pseudo-code template for a single cell of the nested loop constructed for each user-specified stage. 
Each cell is instantiated according to its configuration file, including operation type, source node set, destination node set, time window, and skip/break constraints. 
The compiler then automatically assembles these cells into a full nested-loop structure, respecting the logical order of stages and applying hardware-aware optimizations.
}

\begin{algorithm}[h]
\caption{\textcolor{black}{Generic compiled pseudo-code for one stage of a user-specified AML pattern (\texttt{form\_nested\_loop}).}}
\label{alg:nestedloopcell}
\begin{algorithmic}[1]

\State \textcolor{blue}{// Stage configuration}
\State \textcolor{blue}{// cfg.op $\in$ \{for\_all, union, intersect, differentiate\}}
\State \textcolor{blue}{// cfg.src: input node set (graph-based or prior stage variable)}
\State \textcolor{blue}{// cfg.dst\_var: output node-set variable for this stage}
\State \textcolor{blue}{// cfg.constraints: \{skip\_if, break\_if (incl.\ time window)\}}

\vspace{0.15cm}

\State $edge\_start \gets$ \texttt{Find\_Starting\_Edge}$(t, cfg.src)$

\For{\textbf{each} $e$ \textbf{in} \texttt{IterateEdges}$(cfg.src, edge\_start, cfg.op)$}
    \State \textcolor{blue}{// iterate according to op}

    \If{$e$ satisfies $cfg.constraints.break\_if$}
        \State \textcolor{blue}{// including time-window overflow}
        \State \textbf{break}
    \EndIf

    \If{$e$ satisfies $cfg.constraints.skip\_if$}
        \State \textbf{continue}
    \EndIf

    \State $n \gets$ \texttt{ExtractNode}$(e)$
    \State \textcolor{blue}{// output of each stage is nodes, not edges}

    \State \texttt{cfg.dst\_var.add}$(n)$
    \State \textcolor{blue}{// surviving nodes feed the next stage of the nested loop}

\EndFor

\vspace{0.1cm}
\If{\texttt{is\_final\_stage}}
    \State \texttt{Postprocess}(cfg.dst\_var)
    \State \textcolor{blue}{// final stage result assembly}
\EndIf

\end{algorithmic}
\end{algorithm}

\textcolor{black}{
Beyond direct translation, \THISWORK\ treats all optimization decisions holistically. 
Using information such as neighborhood size, expected cardinality of intermediate sets, and temporal-window selectivity, the compiler decides how each stage should be evaluated, which set operations should be reordered, and whether the stage is better executed on CPU, GPU, or as a hybrid pipeline. 
Optimizations include GPU load balancing, memory-coalesced neighbor iteration, ordering set operations based on estimated cost, and avoiding unnecessary materialization through on-demand propagation. 
Because these optimizations are performed automatically, AML analysts can write patterns at a high level without needing expertise in parallel programming or hardware management.
This compiler-based approach thus enables \THISWORK\ to support a wide range of AML patterns—exact, fuzzy, or hybrid—within one unified abstraction. 
}

\textbf{Input specification and transformation process.}
Each stage in the input specification defines five key components: 
\begin{enumerate}[leftmargin=*]
\item Input operands (such as \textbf{\texttt{N0.out\_neigh}} or \textbf{\texttt{N1.in\_neigh}}),

\item Operation types (\textbf{\texttt{for\_all}} or intersection)

\item Optional temporal constraints (like \textbf{\texttt{Find\_Start(t-$\delta$)}} for time window boundaries)
\item Filtering conditions for skipping nodes or early termination based on timestamp thresholds, and
\item Output variables containing intermediate results passed to subsequent stages.
\end{enumerate}
The compiler parses these specifications, validates operand dependencies between stages to ensure logical consistency, and maps the abstract operations onto highly optimized code templates specifically designed for graphs exhibiting power-law degree distributions common in financial networks.

The generated code templates incorporate sophisticated optimizations tailored for real-world financial transaction graphs.
These optimizations include memory access patterns specifically designed for skewed degree distributions, intelligent workload balancing across CPU threads and GPU warps, efficient set intersection algorithms using degree-based ordering to minimize comparison operations, and overlapping CPU-GPU execution strategies that pipeline data transfer with mining operations across temporal windows.
The compiler automatically selects and applies these optimizations based on the pattern structure and target hardware architecture.
Next, we discuss examples of two patterns to showcase how the compiler outputs high-performance code.

\begin{figure*}[t]
    \centering
    \includegraphics[width=0.98\textwidth]{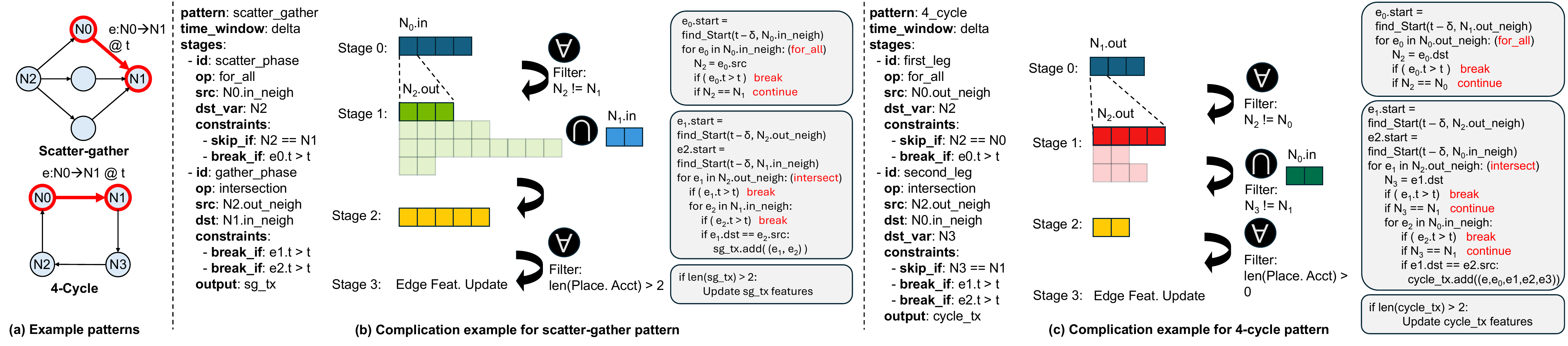}
    \caption{Compiler-generated pseudo-code for scatter-gather and 4-cycle pattern mining.}
    \label{fig:compile_code_example}
\end{figure*}
\textbf{Scatter-gather pattern compilation example.}
As shown in Fig.~\ref{fig:compile_code_example}(b), the input specification defines this pattern through two logical stages: a scatter phase and a gather phase.
The scatter phase uses a \textbf{\texttt{for\_all}} operation to enumerate neighbors from \textbf{\texttt{N0.in\_neigh}}, storing results in variable \textbf{\texttt{N2}}, with constraints to skip cases where \textbf{\texttt{N2 == N1}} and break if the edge timestamp \textbf{\texttt{e0.t}} exceeds the time threshold \textbf{\texttt{t}}.
The gather phase employs an intersection operation between \textbf{\texttt{N2.out\_neigh}} and \textbf{\texttt{N1.in\_neigh}}, incorporating temporal constraints to break if either \textbf{\texttt{e1.t}} or \textbf{\texttt{e2.t}} exceeds the time limit, and outputs results to the \textbf{\texttt{sg\_tx}} variable.

The compiler transforms this specification into optimized kernels that exploit the parallel nature of neighbor enumeration and set intersection operations.
For GPU execution, the compiler generates CUDA kernels that assign individual threads to process different candidate nodes in parallel, utilizing coalesced memory access patterns and shared memory for efficient neighbor list processing.
The temporal constraints are compiled into conditional branches that enable early termination, reducing unnecessary computation when time windows are exceeded.

\textbf{4-Cycle pattern compilation example.}
As illustrated in Fig.~\ref{fig:compile_code_example}(c), the four-node cycle pattern demonstrates the compiler's versatility in handling different topological structures using the same primitive operations. 
The specification defines two stages: a first leg that uses \textbf{\texttt{for\_all}} to traverse from \textbf{\texttt{N0.out\_neigh}} to variable \textbf{\texttt{N2}}, with constraints to skip self-loops (\textbf{\texttt{N2 == N0}}) and respect temporal boundaries, and a second leg that performs an intersection between \textbf{\texttt{N2.out\_neigh}} and \textbf{\texttt{N0.in\_neigh}} to identify closing nodes \textbf{\texttt{N3}}, filtering out cases where \textbf{\texttt{N3 == N1}} and enforcing temporal constraints on both \textbf{\texttt{e1.t}} and \textbf{\texttt{e2.t}}.

The compiler recognizes that cycle detection requires different optimization strategies compared to scatter-gather patterns.
For the cycle pattern, the generated code prioritizes memory locality for sequential neighbor traversals and implements efficient cycle completion checks.
The intersection operation in the second leg is optimized using sorted neighbor lists and binary search techniques, taking advantage of the typically sparse connectivity in financial transaction graphs to minimize intersection complexity.

\textbf{Optimization strategies and code generation.}
The compiler's sophistication lies in its ability to automatically select and apply pattern-specific optimizations without manual intervention.
For scatter-gather, fan, and cycle patterns, the system generates hybrid CPU/GPU code that assigns different traversal stages to the most suitable architecture, thereby achieving additional speedup and leveraging the strengths of both. 
\THISWORK\ maps shallow traversals to the GPU, mitigating load imbalance from skewed workloads while exposing massive parallelism through GPU-friendly primitives for efficient pattern matching. 
A small number of deep traversals are delegated to a CPU post-processing stage, which benefits from the CPU’s high frequency and latency-optimized memory hierarchy. 
Finally, \THISWORK\ produces fully integrated CPU–GPU code and manages intermediate data movement across the two architectures transparently.

This design philosophy enables flexibility in pattern expression and optimization.
Many common AML patterns, including fan-in, fan-out, multi-hop cycles, and their variants, can be expressed as combinations of the same fundamental primitives. 
Even minor changes in operand selection or stage ordering within the input specification can lead the compiler to generate entirely different optimized execution plans, automatically adapting to the structural characteristics of each pattern without requiring manual code modification.
This adaptability, combined with the high-level declarative interface, positions the compiler as a powerful tool for both AML researchers developing new detection algorithms and practitioners deploying production-scale financial crime detection systems.

\section{Evaluation Methodology} \label{section:methodology}
\subsection{Benchmarked dataset}
We use the state-of-the-art synthetic money laundering transaction dataset generated by IBM research~\cite{altman2024realistic}. 
The dataset (Table~\ref{table:dataset}) consists of a set of realistic, standardized AML datasets that have also been evaluated in prior works~\cite{blanuvsa2024graph}. 

\begin{table}[t]
\vspace{-7.5pt}
\caption{IBM AML Datasets~\cite{altman2024realistic}.} 
\vspace{-7.5pt}
\centering 
\scriptsize
\begin{tabular}{c | c | c } 
\hline 
\textbf{Category} & \textbf{Num. of Vertex} & \textbf{Num. of Edges}
\\
\hline 
\hline 
LI-Small & 705,907 & 6,924,055
\\
LI-Medium & 2,032,095 & 31,251,483
\\
LI-Large & 2,070,980 & 176,066,557
\\
HI-Small & 515,088 & 5,078,345
\\
HI-Medium & 2,077,023 & 31,898,238
\\
HI-Large & 2,116,168 & 179,702,229
\\
\hline

\hline 
\end{tabular}
\label{table:dataset} 
\end{table}

\subsection{State-of-the-art Baselines} \label{subsection:baselines}
We compare against two state-of-the-art baselines:
\begin{itemize}[leftmargin=*]
    \item GFP~\cite{blanuvsa2024graph} from IBM, which mines the money laundering patterns, augments edge features, and uses a gradient boost-based classifier downstream, similar to our pipeline.
    \item FraudGT~\cite{lin2024fraudgt}, which uses a graph transformer network to detect money laundering instances.
\end{itemize}

\subsection{Hardware Platform Configuration} \label{subsection:simulation_infrastructure}
We run CPU baselines on a dual-socket server with two Intel Xeon Platinum 8380 processors, each with 40 physical cores (80 SMT threads) and 8 memory channels with a total of 1TB main memory. 
We use up to four NVIDIA A40 GPUs to evaluate our design, each with 48GB GDDR6 memory.

\section{Evaluation Results} \label{section:results}

\begin{figure*}
    \centering
    \includegraphics[width=0.98\textwidth]{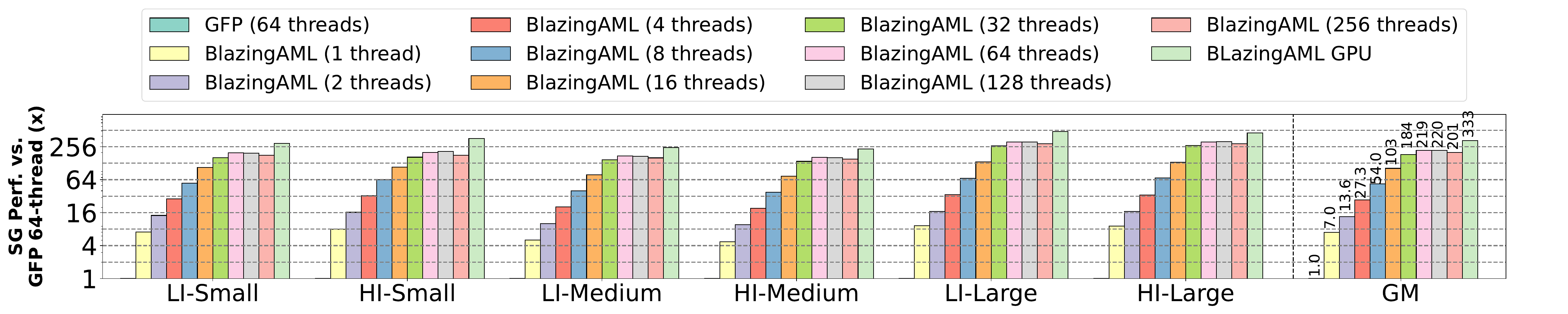}
    \caption{\textbf{\THISWORK\ \texttt{Scatter-Gather} pattern mining end-to-end throughput normalized to GFP~\cite{blanuvsa2024graph}.
    }}
    \label{fig:perf_sg}
\end{figure*}

\begin{figure*}
    \centering
    \includegraphics[width=0.98\textwidth]{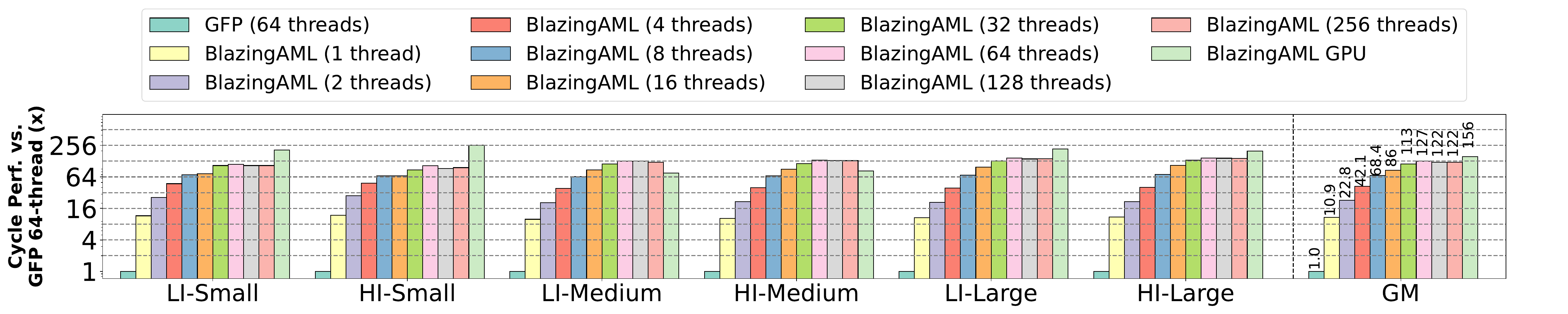}
    \caption{\textbf{\THISWORK\ \texttt{Cycle} pattern mining end-to-end throughput normalized to GFP~\cite{blanuvsa2024graph}.
    }}
    \label{fig:perf_cycle}
\end{figure*}



\subsection{F1 Score Comparison}
Table~\ref{table:f1} shows the F1 score in classifying transactions into money-laundering or normal transactions. 
The features extracted from each transaction (edges) are through pattern matching mined from the graph, \textit{, i.e.}, the value of each feature corresponds to the number of instances of each pattern the transaction participates in. 
Then, the additional features are concatenated and passed to XGB to generate hidden dimensions for training for the first 80\% of the timestamped transactions. 
The inference testing is conducted on the last 20\% of the transactions to infer whether a transaction is fraudulent. 
Table~\ref{table:f1} clearly shows that the addition of structural features (Fan → Degree → Cycle → Scatter-Gather) correlates with consistent performance improvements across all datasets, with the full combination (Fan+Degree+Cycle+SG) achieving peak scores in every case. 
This suggests cumulative benefits from incorporating local node characteristics (Fan/Degree) and global structural patterns (Cycle/SG). 
Notably, the HI (High-illicit) datasets substantially outperform their LI (Low-illicit) counterparts (e.g., HI-Large:58.1 vs LI-Large:17.8), indicating feature effectiveness scales with the density in fraudulent transactions. 
The clear performance hierarchy (Scatter-Gather > Cycle > Degree > Fan) establishes feature contribution weights that guide future model development.

\begin{table}[ht]
\centering
\scriptsize
\caption{\textbf{F1 scores of different features on datasets. \texttt{XGB} is the baseline with source and destination account ID. Additional features include the number of instances of each pattern (\texttt{Fan}, \texttt{Degree}, \texttt{Cycle}, and \texttt{Scatter-Gather}) each transaction participates in.} }
\label{table:f1xgb}
\begin{tabularx}{0.50\textwidth}{Y|Y|Y|Y|Y|Y}
\midrule
\rowcolor{white}
\textbf{Dataset} & \textbf{XGB Only} & \textbf{Fan} & \textbf{Fan+Degree} & \textbf{Fan+Degree+\newline Cycle} & \textbf{Fan+Degree+\newline Cycle+SG}  \\
\hline
\hline
LI-Small & \cellcolor{white} 10.1 & \cellcolor{white} 8.1 & \cellcolor{white} 10.3 & \cellcolor{white} 12.7 & \cellcolor{ForestGreen} 18.4  \\
HI-Small & \cellcolor{white} 11.1 & \cellcolor{white} 23.2 & \cellcolor{white} 38.4 & \cellcolor{white} 42.6 & \cellcolor{ForestGreen} 46.6  \\
LI-Med & \cellcolor{white} 3.1 & \cellcolor{white} 4.0 & \cellcolor{white} 11.0 & \cellcolor{white} 14.0 & \cellcolor{ForestGreen} 21.4  \\
HI-Med & \cellcolor{white} 10.4 & \cellcolor{white} 29.3 & \cellcolor{white} 47.1 & \cellcolor{white} 50.4 & \cellcolor{ForestGreen} 51.1  \\
LI-Large & \cellcolor{white} 5.4 & \cellcolor{white} 9.1 & \cellcolor{white} 16.0 & \cellcolor{white} 17.7 & \cellcolor{ForestGreen} 17.8  \\
HI-Large & \cellcolor{white} 20.2 & \cellcolor{white} 39.1 & \cellcolor{white} 55.7 & \cellcolor{white} 57.5 & \cellcolor{ForestGreen} 58.1  \\
\hline
\end{tabularx}
\label{table:f1}
\end{table}

\subsection{Mining Performance Comparison}

\begin{figure*}
    \centering
    \includegraphics[width=0.98\textwidth]{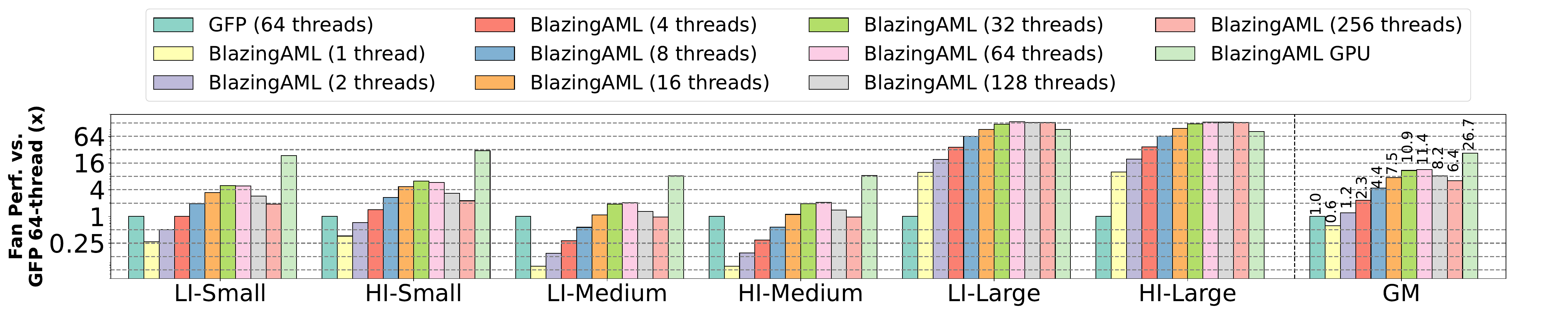}
    \caption{\textbf{\THISWORK\ \texttt{Fan-in} and \texttt{Fan-out} pattern mining combined end-to-end throughput normalized to GFP~\cite{blanuvsa2024graph}.
    }}
    \label{fig:perf_fan}
\end{figure*}

\begin{figure*}
    \centering
    \includegraphics[width=0.98\textwidth]{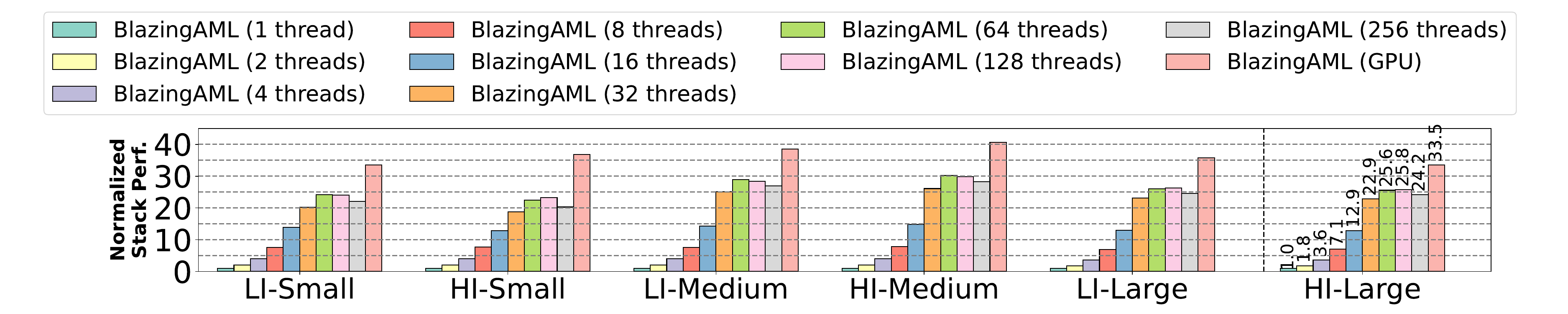}
    \caption{\textbf{\THISWORK\ \texttt{Stack} pattern mining end-to-end throughput normalized to single-thread CPU.
    }}
    \label{fig:perf_stack}
\end{figure*}

\begin{figure*}
    \centering
    \includegraphics[width=0.98\textwidth]{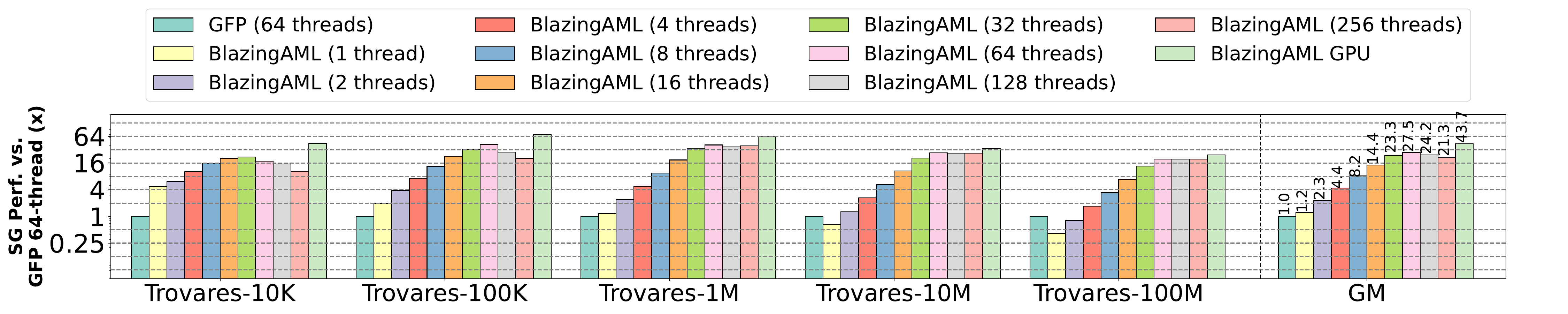}
    \caption{\textbf{Scalability study of \THISWORK\ \texttt{Scatter-Gather} pattern mining throughput normalized to GFP~\cite{blanuvsa2024graph} on Trovares~\cite{trovares22Temporal} 10K -- 100M edge dataset.
    }}
    \label{fig:trovares}
\end{figure*}
Fig.~\ref{fig:perf_sg}, ~\ref{fig:perf_cycle}, ~\ref{fig:perf_fan}, ~\ref{fig:perf_stack} shows the performance study of \THISWORK\ framework vs. GFP library~\cite{blanuvsa2024graph}. 
\THISWORK\ is implemented with OpenMP parallelism and evaluated from 1 to 256 threads, as well as CUDA implementation on a single A40 GPU. 
\textbf{Scatter-Gather Pattern Performance}: Compared to 64-thread GFP library implementation in mining \texttt{Scatter-Gather} pattern, \THISWORK\ already achieves performance comparable to GFP even with CPU single thread, underscoring the performance potentials in legacy python-based libraries.
The mining performance scales almost linearly with thread count up to 64 threads, reaching 219–220$\times$ speedup over GFP (64 threads) depending on dataset.
At 128 and 256 threads, scalability continues, achieving peak improvements of 333$\times$ over GFP on the largest datasets.

\textbf{Cycle Pattern Characteristics}: The \texttt{Cycle} pattern demonstrates high performance advantages, with single-thread performance already achieving 10.9$\times$ speedup over GFP's 64-thread baseline. 
This pattern exhibits a high scaling behavior, reaching 159$\times$ improvement with GPU acceleration. 
The sustained linear scaling up to 64 threads (127$\times$ speedup) followed by a performance plateau at higher thread counts suggests that cycle detection benefits significantly from the optimized graph traversal algorithms.

\textbf{Fan Pattern}: 
For less complex logic such as \texttt{Fan-in} and \texttt{Fan-out}, the GFP library 64-thread achieves comparable performance with \THISWORK\ less than 8 threads. 
\THISWORK\ demonstrates consistent improvement up to 32 threads (11.4$\times$) before experiencing performance degradation at higher thread counts (8.2$\times$ at 128 threads). 
For GPU, the data structure of neighborhood search is further optimized in CUDA to achieve a better speedup for basic patterns such as \texttt{Fan}. 

\textbf{Stack Pattern Baseline Comparison}: 
The \texttt{Stack} pattern evaluation uses a different baseline (1-thread vs. GFP 64-thread), making direct comparison challenging, but reveals excellent parallel scalability up to 64 threads (25.8$\times$) with slight degradation at higher core counts. The GPU achieves 33.5$\times$ improvement, demonstrating consistent acceleration across all evaluated patterns.

The pattern mining results demonstrate both strong parallel scalability and effective load distribution in the generated code across diverse graph mining workloads. 
The results confirm that \THISWORK's domain-specific compiler produces significantly more efficient code than hand-optimized existing baseline implementations.
\THISWORK\ scales to hundreds of cores, though with pattern-specific behaviors that suggest intelligent workload-aware optimization strategies.
The performance benefits hold consistently across all dataset categories (low vs. high interconnectivity, small vs. large), demonstrating the generality of the compiler design.
While the GPU backend provides strong performance, high-core-count CPUs deliver competitive speedups, giving users flexibility in choosing the hardware platform based on the specific pattern mining requirements.

\subsection{Scalability Study}
Figure~\ref{fig:trovares} evaluates the scalability of \THISWORK\ against the baseline GFP implementation using synthetic graphs generated by Trovares~\cite{trovares22Temporal} spanning five orders of magnitude in size, from 10K to 100M edge. 
On the smallest dataset (Trovares-10K), \THISWORK\ achieves a 21.8$\times$ speedup with 32-thread execution. 
On larger datasets, \THISWORK\ achieves a 40.8$\times$ and 27.8$\times$ speedup in Trovares-1M and Trovares-10M. 
The average speedup of \THISWORK\ GFP reaches a remarkable 27.5$\times$ speedup using the same 64 threads, indicating excellent parallel scalability of our scatter-gather pattern implementation.
Our multi-threading analysis reveals consistent scaling behavior across different thread counts. 
The performance improvements scale nearly linearly from 1 to 64 threads, with the 64-thread configuration achieving speedups of 19.5$\times$ and 27.5$\times$ on Trovares-10M and 100M.

\THISWORK\ GPU implementation demonstrates a 24.4$\times$ speedup on Trovares-100M compared to the baseline GFP. 
This substantial improvement highlights the effectiveness of our CUDA implementation in exploiting the massive parallelism inherent in scatter-gather operations on large-scale graphs.
The scalability trends clearly favor \THISWORK\ as dataset size increases. 
This behavior is consistent with our design philosophy of optimizing for large-scale graph mining workloads where the overhead of our compilation framework is amortized across substantial computational work.

\subsection{\textcolor{black}{F1 Score of Money-Laundering Predictions}}
\textcolor{black}{
\textbf{Why using F1 score?}
Money laundering is inherently an imbalanced problem since laundering transactions only account for a very small fraction of the total transactions in real-world financial systems.
In such an imbalanced inference setting, the F1 score is widely adopted as the primary evaluation metric~\cite{blanuvsa2024graph, song2024identifyingmoneylaunderingsubgraphs} because it balances precision and recall and prevents models from achieving deceptively high accuracy by simply predicting the majority “non-laundering” class.
}
\textcolor{black}{
To further highlight the imbalance in real AML datasets, Table~\ref{tab:confusion_f1} shows the confusion matrix obtained on the HI-Small dataset after applying all mined features (Fan, Degree, Cycle, and SG) and training the XGB classifier. 
Even in this relatively high–laundering-occurrence dataset, the TN count remains overwhelmingly dominant. 
This concrete example reinforces why the F1 score is the most appropriate metric for evaluating laundering prediction performance in such imbalanced conditions.
}

\begin{table*}[h]
    \centering
    \caption{\textcolor{black}{Confusion matrix of laundering prediction on HI-Small after applying all \THISWORK\ features. The matrix demonstrates the extreme class imbalance even in a high-laundering dataset.}}
    \label{tab:confusion_f1}
    \begin{tabular}{c|c|c}
        \textbf{} & \textbf{Predicted Laundering} & \textbf{Predicted Non-Laundering} \\
        \hline
        \textbf{Actual Laundering} & 558 & 1239 \\
        \textbf{Actual Non-Laundering} & 31 & 1013841 \\
    \end{tabular}
\end{table*}
\begin{figure*}
    \centering
    \includegraphics[width=0.98\textwidth]{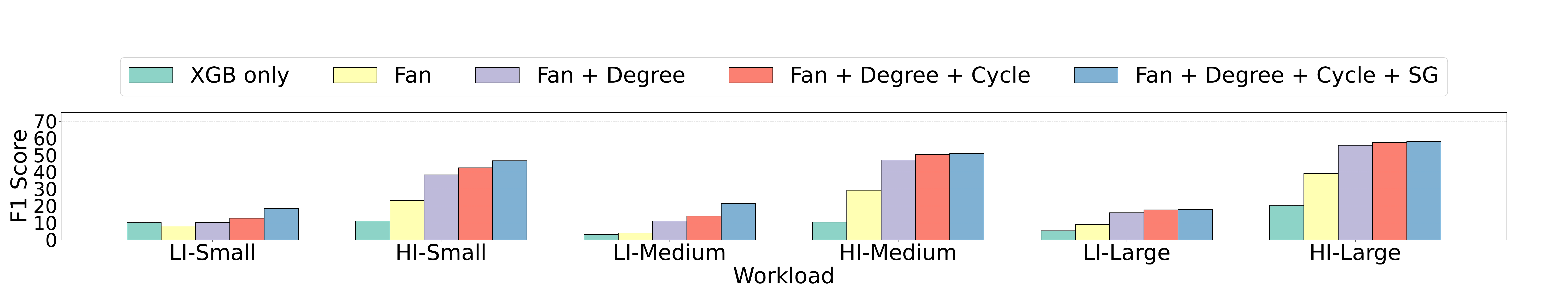}
    \caption{\textcolor{black}{\textbf{F1 score of \THISWORK\ when using different shapes as mining features. The F1 score increases as more features (the number of participating shapes for each transactional edge) are included and then used for training and inference via the XGB Boost library. \THISWORK\ preserves the output quality of the GFP library while providing a faster and more flexible mining framework.}}}
    \label{fig:f1scorewide}
\end{figure*}
\textcolor{black}{
Figure~\ref{fig:f1scorewide} presents the F1 score of predicting whether a transaction is a laundering transaction under different feature configurations.
In the baseline setting, only the raw transactional information (transaction ID, source account, and destination account) is used as features for XGB Boost.
As additional features derived from pattern mining are included—such as the number of Scatter-Gather (SG) structures an edge participates in—the F1 score increases significantly.
This demonstrates that structural mining features contribute strong discriminative power in identifying laundering behaviors.
}

\textcolor{black}{
\THISWORK\ intentionally keeps the set of mining outputs identical to the GFP library to ensure full compatibility, while providing a more efficient and flexible execution engine.
Across all datasets, Degree, Cycle, and SG features consistently lead to meaningful improvements in F1 score, confirming their effectiveness in capturing laundering-related topological signals.
}

\textcolor{black}{
An additional trend observed in Figure~\ref{fig:f1scorewide} is that the HI datasets (with higher laundering occurrence) achieve substantially higher F1 scores than the LI datasets.
This is expected: in datasets with more positive instances, the classifier receives stronger supervision and can learn laundering-related behaviors more reliably.
}

\subsection{Comparison with FraudGT}
\begin{table}[ht]
\centering
\scriptsize
\caption{\textbf{F1 scores of \texttt{FraudGT}~\cite{lin2024fraudgt} and \THISWORK\ across IBM datasets.}}
\begin{tabularx}{0.48\textwidth}{p{0.075\textwidth}|Y|Y|Y|Y|Y|Y}
\midrule
\rowcolor{white}
\textbf{Method} & \textbf{LI-Small} & \textbf{HI-Small} & \textbf{LI-Medium} & \textbf{HI-Medium} & \textbf{LI-Large} & \textbf{HI-Large} \\
\hline
\hline
FraudGT    & \cellcolor{white} 28.6 & \cellcolor{white} 69.6 & \cellcolor{white} \textbf{24.0} & \cellcolor{white} \textbf{62.3} & \cellcolor{white} 11.0 & \cellcolor{white} 54.3 \\
\THISWORK\  & \cellcolor{white} 18.4 & \cellcolor{white} 46.4 & \cellcolor{white} 21.4 & \cellcolor{white} 51.1 & \cellcolor{white} \textbf{17.8} & \cellcolor{white} \textbf{58.1} \\
\hline
\end{tabularx}
\label{table:f1_fraudgt}
\end{table}
\begin{figure}
    \centering
    \includegraphics[width=0.48\textwidth]{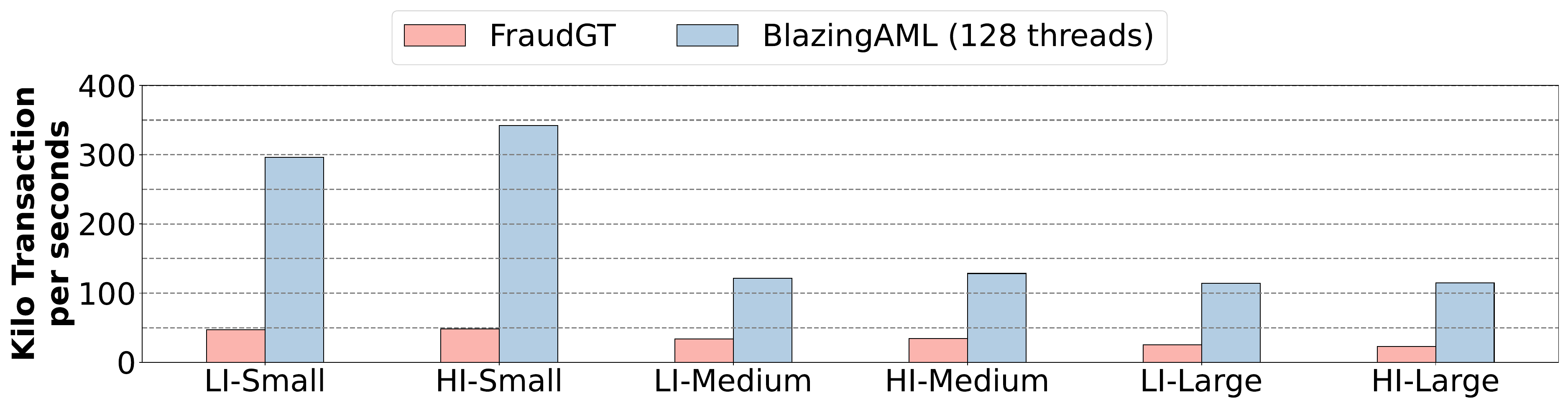}
    \caption{\textbf{Performance study of \THISWORK\ compared with FraudGT. 
    \THISWORK\ processes 4.9$\times$ higher number of edges per second on average. 
    }}
    \label{fig:fraudgt}
\end{figure}

FraudGT~\cite{lin2024fraudgt} uses a graph transformer model and achieves state-of-the-art performance in detecting fraudulent activities, also demonstrating high throughput compared to existing methods. 
We compare \THISWORK\ and FraudGT on their achieved F1 score and mining throughput in detecting money laundering patterns. 
Tab.~\ref{table:f1_fraudgt} shows the F1 score of the two mining framework. 
Note that we have verified that \THISWORK\ achieves the same feature mining output as GFP~\cite{blanuvsa2024graph}. 
Therefore, the F1 score difference roots from the ML framework difference (feature extension + XGB in ~\cite{blanuvsa2024graph} and Transformer-based model in ~\cite{lin2024fraudgt}). 
Fig.~\ref{fig:fraudgt} compares the mining throughput of \THISWORK\ with FraudGT. 
On average, \THISWORK\ 128-thread implementaion achieves 4.9$\times$ higher throughput, which corroborates with the intuition that feature mining+XGB achieving a much more efficient solution than Transformer-based models. 

The performance gap is particularly pronounced on larger datasets, where \THISWORK\ processes between 300-400 thousand transactions per second compared to FraudGT's 50-100 thousand transactions per second. 
On average across all configurations, \THISWORK's 128-thread implementation achieves a 4.9$\times$ higher throughput than FraudGT.
This substantial performance improvement can be attributed to the fundamental algorithmic differences between the two approaches. 
This design choice proves particularly effective for fraud detection scenarios where rapid transaction processing is critical, as our feature extraction and gradient boosting approach scales more efficiently than the quadratic complexity inherent in Transformer architectures.

\section{Conclusion} \label{section:conclusion}
This paper presented \THISWORK, a scalable anti-money laundering (AML) system that advances the way financial institutions design and deploy pattern detection algorithms.
At the core of our approach is a \textit{multi-stage specification technique} that captures both the structural and temporal complexities of money laundering schemes, moving beyond the limitations of traditional rigid pattern-matching techniques.
A domain-specific compiler bridges AML expertise and high performance deployment, automatically generating optimized implementations without requiring low-level programming knowledge.
Our evaluation demonstrates that \THISWORK\ achieves substantial speedups on both CPUs and GPUs while preserving high detection accuracy, making sophisticated pattern mining both practical and scalable.

\clearpage
\balance

\bibliographystyle{ACM-Reference-Format}
\bibliography{99_ref.bib}

\end{document}